\documentclass[12pt, a4paper]{amsart}
\usepackage{amsmath, bm}
\usepackage{graphicx,psfrag,epsf}
\usepackage{enumerate}
\usepackage[authoryear]{natbib}
\usepackage{url} 
\usepackage{amsaddr}

\usepackage{subcaption}
\usepackage{amsfonts, bm, bbm}
\usepackage{amssymb}
\usepackage[makeroom]{cancel}
\usepackage{hhline}
\usepackage{caption}
\usepackage[labelformat=simple]{subcaption}

\usepackage{enumitem}

\usepackage{algorithm}
\usepackage{algpseudocode}
\usepackage{multirow}
\usepackage{arydshln}

\newdimen\figrasterwd
\figrasterwd\textwidth

\usepackage{color}
\usepackage{booktabs}

\makeatletter
\newcommand{\pushright}[1]{\ifmeasuring@#1\else\omit\hfill$\displaystyle#1$\fi\ignorespaces}
\newcommand{\pushleft}[1]{\ifmeasuring@#1\else\omit$\displaystyle#1$\hfill\fi\ignorespaces}
\makeatother

\newcommand\independent{\protect\mathpalette{\protect\independenT}{\perp}}
\def\independenT#1#2{\mathrel{\rlap{$#1#2$}\mkern2mu{#1#2}}}

\definecolor{brown}{rgb}{0.8, 0.33, 0.1}

\def\bx {\bm x}

\def\bX {\bm X}

\def\bbI {\mathbf I}

\def\bbeta {\bm \beta}

\def\bmu {\bm \mu}

\def\bbOmega {\mathbf \Omega}

\def\P {\text{P}}

\def\true {\text{true}}
\def\T {\mathcal{T}}
\def\M {\mathcal{M}}
\def\C {\text{C}}
\def\P {\text{P}}

\def\E {\text{E}}

\def\N {\text{N}}

\def\Ber {\text{Ber}}
\def\IG {\text{IG}}

\def\undertilde#1{\mathord{\vtop{\ialign{##\crcr
$\hfil\displaystyle{#1}\hfil$\crcr\noalign{\kern1.5pt\nointerlineskip}
$\hfil\tilde{}\hfil$\crcr\noalign{\kern1.5pt}}}}}

\makeatletter
\def\paragraph{\@startsection{paragraph}{4}%
  \z@{1ex \@plus1ex \@minus.2ex}{-\fontdimen2\font}%
  {\normalfont\itshape}}
\makeatother

\begin{document}

\def\spacingset#1{\renewcommand{\baselinestretch}%
{#1}\small\normalsize} \spacingset{1}


\title[]{Incorporating External Data into the Analysis of Clinical Trials via Bayesian Additive Regression Trees}

\author[T. Zhou]{Tianjian Zhou$^1$}
\address{Department of Statistics, Colorado State University \\}
\author[Y. Ji]{Yuan Ji$^2$}
\address{Department of Public Health Sciences, University of Chicago}
\email{$^1$tianjian.zhou@colostate.edu, $^2$yji@health.bsd.uchicago.edu}


\begin{abstract}
Most clinical trials involve the comparison of a new treatment to a control arm (e.g., the standard of care) and the estimation of a treatment effect.
External data, including historical clinical trial data and real-world observational data, are commonly available for the control arm.
Borrowing information from external data holds the promise of improving the estimation of relevant parameters and increasing the power of detecting a treatment effect if it exists. 
In this paper, we propose to use  Bayesian additive regression trees (BART) for incorporating external data into the analysis of clinical trials, with a specific goal of estimating the conditional or population average treatment effect.
BART naturally adjusts for patient-level covariates and captures potentially heterogeneous treatment effects across different data sources, achieving flexible borrowing.
Simulation studies demonstrate that BART compares favorably to a hierarchical linear model and a normal-normal hierarchical model.
We illustrate the proposed method with an acupuncture trial.
\end{abstract}

\spacingset{1.45}

\maketitle

\section{Introduction}
\label{sec:intro}

Most clinical trials involve the comparison of a new treatment to a control arm (e.g., the standard of care) and the estimation of a treatment effect.
While the investigational drug is new, \emph{external data} are usually available for the control arm. Here,
external data can be \emph{historical data} from past clinical trials \citep{viele2014use}, or \emph{real-world data} (RWD) routinely gathered from a variety of sources other than traditional clinical trials \citep{fda2018}.
Examples of RWD include data derived from electronic health records, and medical claims and billing data.
Incorporation of external data into clinical trials holds the promise of improving the estimation of relevant parameters, e.g., reducing the mean squared errors of the parameter estimates and increasing the power of detecting a treatment effect if it exists. 
With additional information about the control arm, more resources can be devoted to the new treatment, and the required total sample size may be reduced.
In the ideal situation, external data can be used to synthesize an external control arm, and the clinical trial can be conducted in a single-arm fashion \citep{goring2019characteristics}.

A variety of statistical methods have been developed for borrowing information from historical trial data, with the majority of these methods being Bayesian.
See, for example, \cite{viele2014use} or \cite{van2018including} for a comprehensive review.
Roughly speaking, methods for historical borrowing can be categorized into two types. 
The first type of methods constructs an informative prior for the parameters in the concurrent control arm based on the (down-weighted) likelihood of the historical data. 
Examples of these methods include the power prior approach \citep{ibrahim2000power, ibrahim2015power}, the modified power prior approach \citep{neuenschwander2009note},  and the commensurate prior approach \citep{hobbs2011hierarchical, hobbs2012commensurate}.
The second type of methods are based on hierarchical modeling, which assumes the concurrent control arm parameters and the historical control parameters are random samples from a common population distribution.
As a result, parameter estimates for the concurrent and historical controls are shrunk towards an overall mean.
Examples of hierarchical modeling approaches include \cite{neuenschwander2010summarizing},  \cite{schmidli2014robust}, \cite{kaizer2018bayesian}, \cite{lewis2019borrowing}, and \cite{rover2020dynamically}.
The two types of methods are closely related, and there is no sharp distinction between them (see, e.g., \citealp{chen2006relationship}). 
To achieve reliable information borrowing, the historical data should satisfy a series of conditions \citep{pocock1976combination}. For example, the historical control group should have been part of a recent clinical study which contained the same requirements for patient eligibility, and the distributions of important patient characteristics in the group should be comparable with those in the current trial.
The use of historical information in medical device trials was discussed by \cite{fda2010} .

Recently, the increasing accessibility of routinely collected health care data has led to interests in the use of RWD for assessing treatment effects in clinical trials 
\citep{sherman2016real, khozin2017real, corrigan2018real, fda2018}.
It is expected that the patient characteristics in the RWD are different from those in the clinical trial. Therefore, techniques in the causal inference literature, such as propensity score weighting or matching, are commonly employed to adjust for such discrepancy.
Examples of recent works on RWD include \cite{ventz2019design} and \cite{carrigan2019using}.

In this paper, we propose to use Bayesian additive regression trees (BART, \citealp{chipman2010bart}) for incorporating external data into the analysis of clinical trials.
Our approach aims to tackle two issues regarding borrowing external data:
\begin{enumerate}
\item The distribution of patient characteristics in the current clinical trial may be different from that in the external data, and
\item Even if the patient-level covariates have been adjusted, there might still be unmeasured confounders which can contribute to heterogeneity in patient outcomes across data sources.
\end{enumerate}
The first issue has been the focus of much of the RWD literature and is also a main problem in causal inference, while the second issue has been the focus of the historical borrowing literature.
By modeling the relationship between patient outcome, covariates, and an indicator of data source using BART, the proposed approach takes into account both issues.
As seen in \cite{hill2011bayesian}, BART is able to detect interactions and nonlinearities thus can readily identify heterogeneous treatment effects across the covariate space and data sources. On the other hand, for areas of the covariate space in which patient outcomes are relatively homogeneous, BART pools information across data sources to produce more precise estimates (Section \ref{sec:model}).
These features allow BART to adaptively incorporate external information into the analysis of clinical trials.

BART has many attractive features.
The implementation of BART has been made easy by the R package \texttt{BART} \citep{sparapani2021nonparametric}, which supports continuous, binary, categorical, and time-to-event outcomes and can also perform variable selection in the presence of a large number of covariates \citep{linero2018bayesian}.
A stream of work has demonstrated the promising theoretical properties of BART (e.g., \citealp{linero2018bayesian2} and \citealp{rovckova2020posterior}).
Lastly, BART has been used in many applications, such as causal inference \citep{hill2011bayesian}, survival analysis \citep{sparapani2016nonparametric}, subgroup finding \citep{sivaganesan2017subgroup}, and missing data \citep{zhou2020semiparametric}, with proven good performance.
A recent review of BART can be found in \cite{hill2020bayesian}.

The remainder of this paper is organized as follows. 
In Section \ref{sec:method}, we describe the proposed methodology, including an introduction of the treatment effect estimation problem in clinical trials.
In Section \ref{sec:simulation}, we present simulation studies and comparisons with a hierarchical linear model (HLM), a normal-normal hierarchical model (NNHM), and versions of these models that ignore external data.
In Section \ref{sec:real_data}, we illustrate the application of our method with an acupuncture trial.
Finally, in Section \ref{sec:discussion}, we conclude with a discussion.

\section{Methodology}
\label{sec:method}

\subsection{Treatment Effect Estimation in Clinical Trials}
\label{sec:def_ate}

Suppose that in addition to the current clinical trial data, we have access to supplemental data from $J$ external data sources (which can be historical trial datasets or real-world datasets).
Let $i$ index an individual in the current trial, a historical study, or a real-world study. Let $S_i$ be an indicator for data source, where $S_i = 0$ indicates that individual $i$ is from the current trial, and $S_i \in \{ 1, 2, \ldots, J \}$ represents that the individual is from an external dataset.
Denote by $T_i = 1$ (or $0$) an assignment of individual $i$ to treatment (or control).
Without loss of generality, assume that all individuals in the external data received the control, while patients in the current trial are assigned to both the treatment and control arms.
Next, following the notation in the causal inference literature (e.g. \citealt{rubin1974estimating}), 
we denote by $Y_i(1)$ or $Y_i(0)$ the potential outcomes that would have been observed for the same individual $i$ if the individual received treatment or control, respectively. 
In practice, $Y_i(1)$ and $Y_i(0)$ cannot be observed simultaneously, and we let $Y_i = Y_i(1)T_i + Y_i(0) (1 - T_i)$ denote the observed outcome. 
At this moment, we consider continuous outcomes and focus on the treatment effect defined by the difference between the potential outcomes, $Y_i(1) - Y_i(0)$. The idea can be easily extended to binary or time-to-event outcomes. 
Finally, denote by $\bX_i = (X_{i1}, \ldots, X_{iQ})^{\top}$ the $Q$ patient-level covariates, which are assumed to be measured in both the current trial and external data.

The primary goal of many clinical trials is to estimate the \emph{average treatment effect} (ATE), which may be defined over the sample or the population. Common estimands include the conditional ATE (CATE) and the population ATE (PATE), defined as
\begin{align*}
\Delta_{\C} &= \frac{1}{N_{\text{trial}}} \sum_{i: S_i = 0}  \E[Y_i(1) -  Y_i(0) \mid \bX_i, S_i = 0], \quad \text{and} \\
\Delta_{\P} &= \E[Y_i(1) -  Y_i(0) \mid S_i = 0] = \E\big[\E[Y_i(1) -  Y_i(0) \mid \bX_i, S_i = 0] \big],
\end{align*}
respectively. Here, $N_{\text{trial}} = \sum_i \mathbbm{1}(S_i = 0)$ denotes the number of individuals in the current trial.
See, for example, \cite{imbens2004nonparametric} or \cite{hill2011bayesian} for a discussion of different types of estimands.
In this paper, we focus on the estimation of the CATE but will also discuss how to estimate the PATE.
For simplicity, we drop the index $i$ whenever needed.

The distribution of $[Y(t) \mid \bX, S]$ is known as the \emph{response surface}. Under the \emph{unconfoundedness} condition that $[Y(0), Y(1) \independent T \mid \bX, S = 0]$ (which is satisfied for randomized or conditionally randomized trials), we have
\begin{align*}
\E[Y(t) \mid \bX, S = 0] &= \E[Y(t) \mid T = t, \bX, S = 0]  \nonumber\\
&= \E[Y \mid T = t, \bX, S = 0].
\end{align*}
As a result, 
\begin{multline*}
\E[Y(1) - Y(0) \mid \bX, S = 0] 
= \\
\E[Y \mid T = 1, \bX, S = 0] - \E[Y \mid T = 0, \bX, S = 0] \triangleq \delta(\bX),
\end{multline*}
and 
\begin{align}
\Delta_{\C} = \frac{1}{N_{\text{trial}}} \sum_{i: S_i = 0} \delta(\bX_i), \qquad
\Delta_{\P} = \int \delta(\bx) p(\bx \mid S = 0) d \bx,
\label{eq:CATE2}
\end{align}
where $p(\bx \mid S = 0)$ is the distribution of patient characteristics over the trial population.
The quantity $\delta(\bx)$ is referred to as the \emph{conditional treatment effect}.

Incorporation of external data could ideally improve the estimation of $\E[Y \mid T = 0, \bX, S = 0]$.
In the ideal case, the control outcome $Y$ and data source indicator $S$ are  \emph{conditionally independent} given the covariates $\bX$, meaning $[Y \independent S \mid T = 0, \bX]$, hence $[Y \mid T = 0, \bX, S = s ] \stackrel{d}{=} [Y \mid T = 0, \bX]$, where $\stackrel{d}{=}$ denotes equality in distribution.
Under conditional independence, the current and external control data can be pooled together to estimate $\E[Y \mid T = 0, \bX]$.
In practice, conditional independence may not hold due to unmeasured confounding covariates.
Our objective is to capture the discrepancy in $[Y \mid T = 0, \bX]$ across data sources by modeling the relationship among $Y$, $\bX$, and $S$ for the control data.
Details on the model specification are presented next in Section \ref{sec:model}.

\subsection{Model Specification via BART}
\label{sec:model}

We specify probability models separately for the control and treatment groups in very general forms,
\begin{align}
[Y \mid T = 0, \bX = \bx, S = s] &= f_0 (\bx, s) + \varepsilon_{0}, \quad \text{and} \label{eq:model_ctl}\\
[Y \mid T = 1, \bX = \bx, S = 0] &= f_1 (\bx) + \varepsilon_{1}, \nonumber
\end{align}
where $\varepsilon_{0} \sim \N(0, \sigma_0^2)$ and $\varepsilon_{1} \sim \N(0, \sigma_1^2)$ are Gaussian errors.
Therefore, 
\begin{align}
\delta(\bx) = f_1(\bx) - f_0(\bx, 0).
\label{eq:CATE3}
\end{align}
In principle, one may use any method that flexibly estimates $f_0$ and $f_1$. Here we choose BART for its many attractive features discussed in Section \ref{sec:intro}.

\begin{figure}[h!]
\begin{center}
\begin{subfigure}[t]{\textwidth}
\centering
\includegraphics[height=6.2cm]{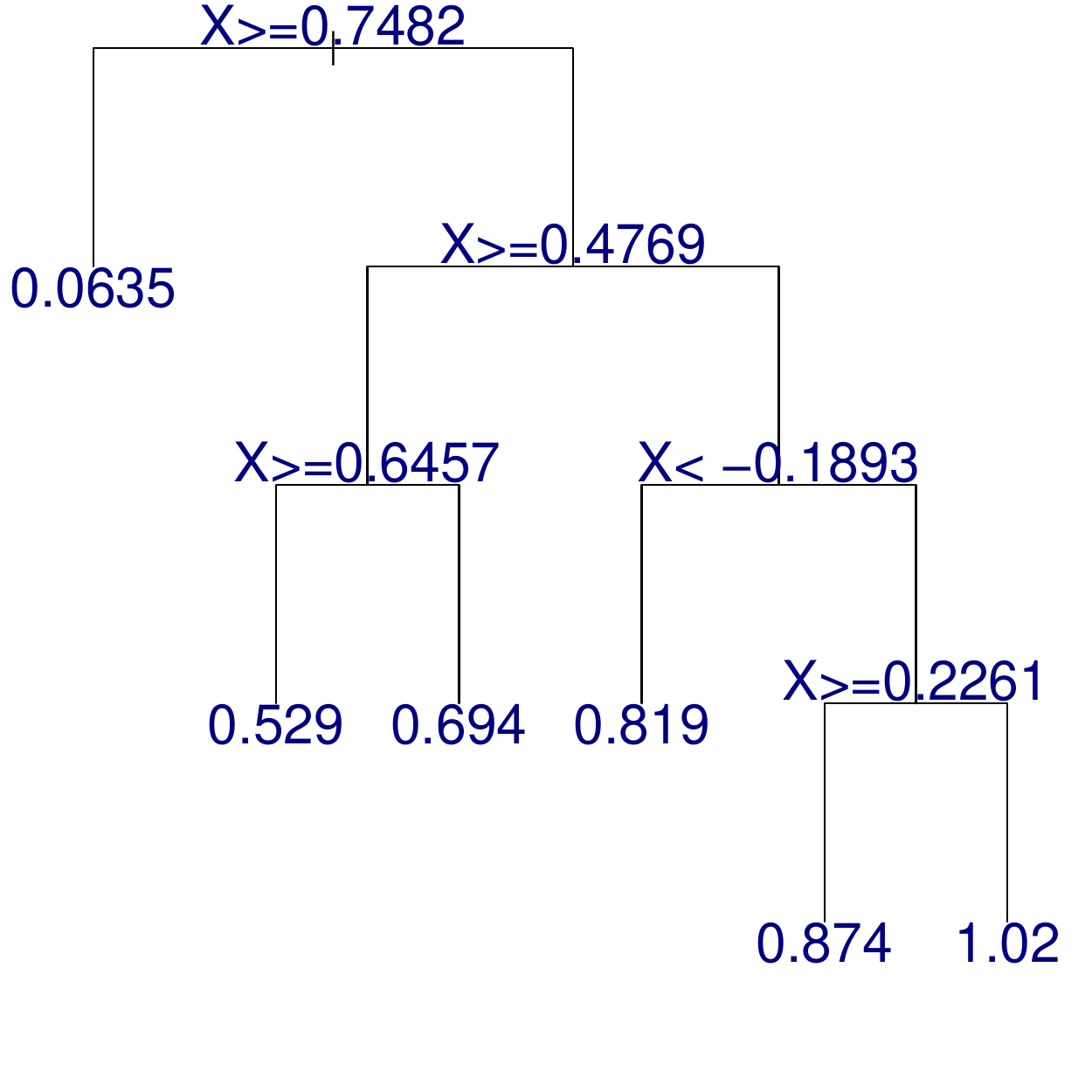}
\includegraphics[height=6.2cm]{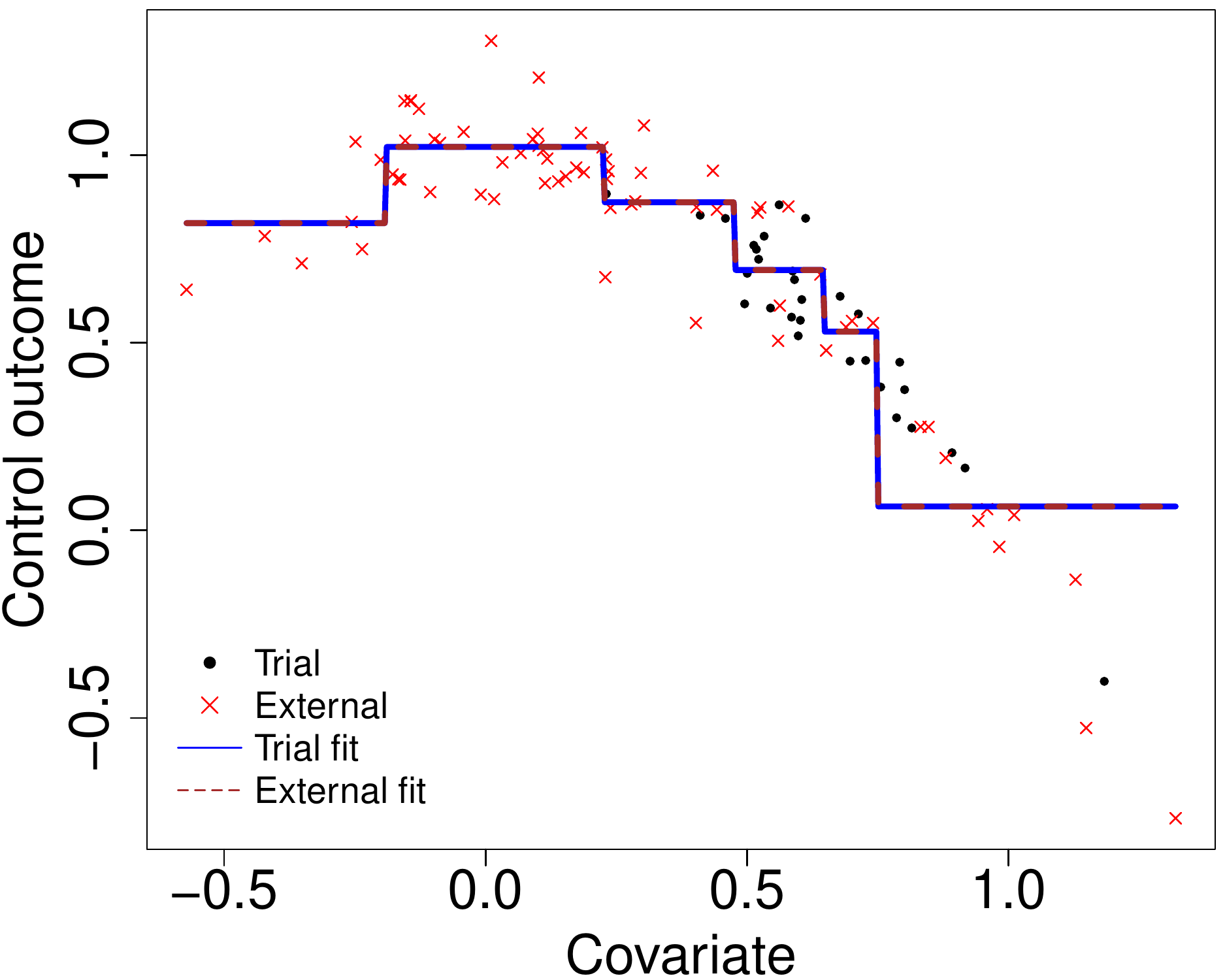}
\caption{Example 1: control outcome is conditionally independent of data source}
\end{subfigure} \\ \vspace{1cm}
\begin{subfigure}[t]{\textwidth}
\centering
\includegraphics[height=6.2cm]{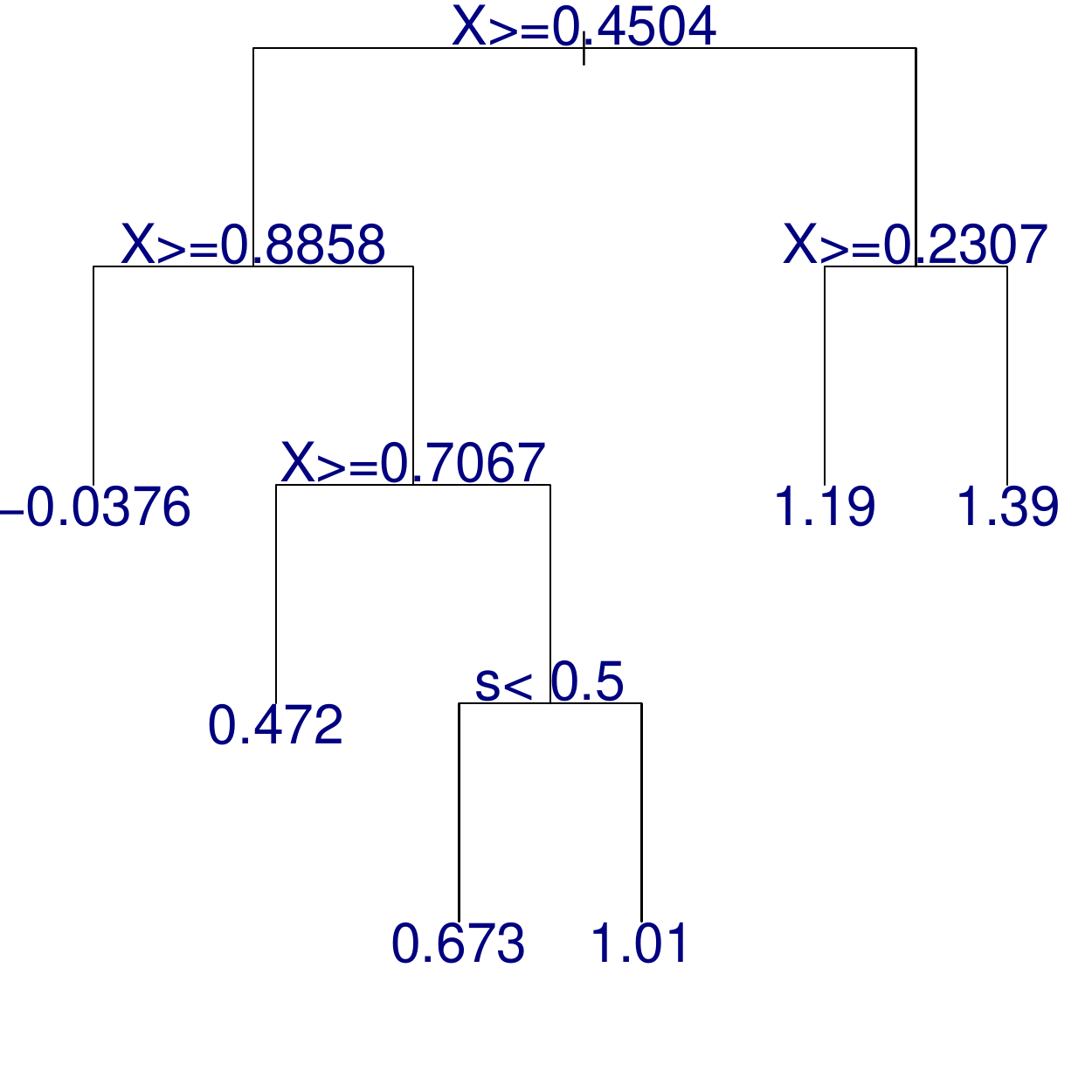}
\includegraphics[height=6.2cm]{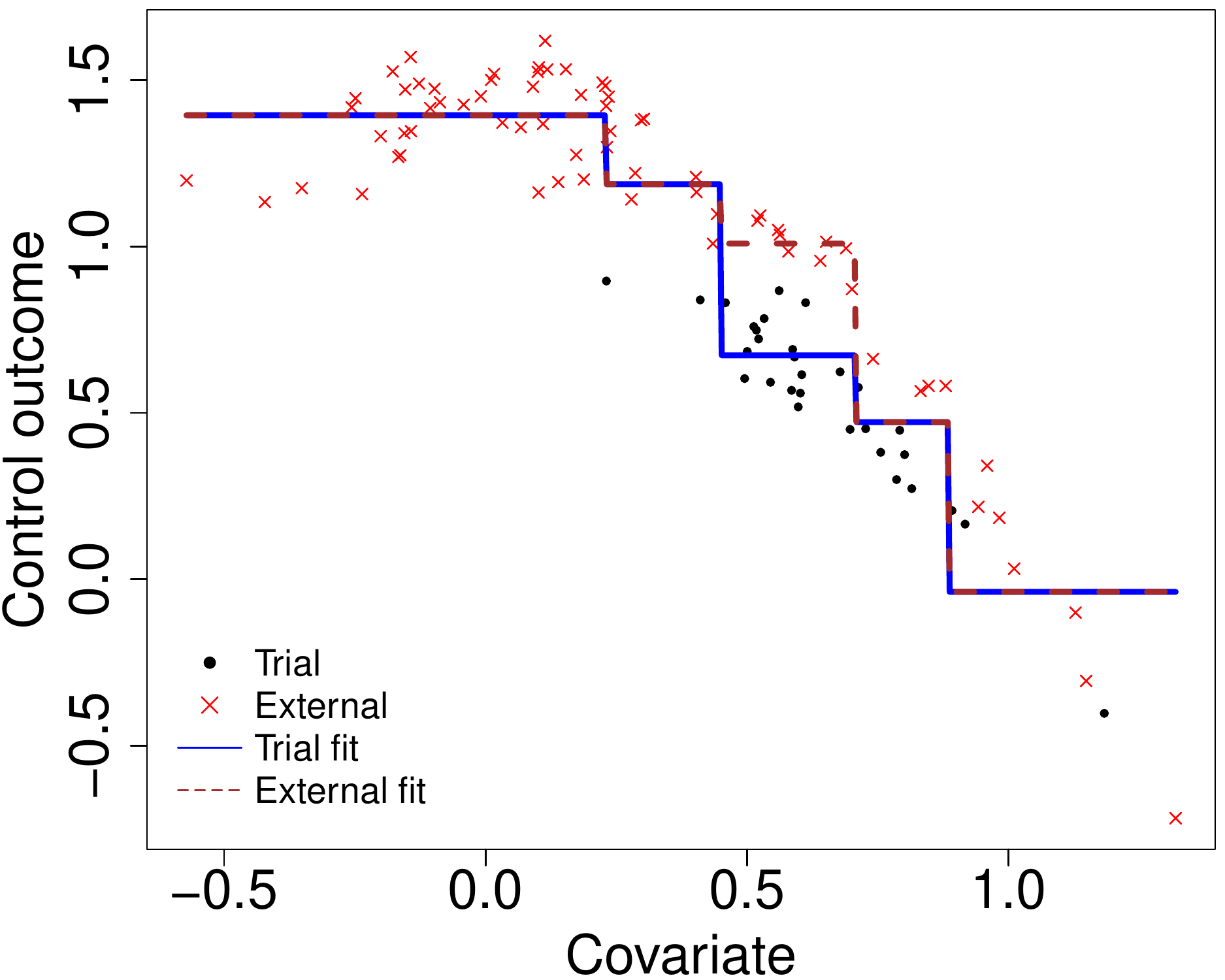}
\caption{Example 2: control outcome is not conditionally independent of data source}
\end{subfigure}
\end{center}
\caption{Single tree fits to two sets of simulated data. In the conditionally independent case, the tree model automatically pools the trial and external control data. When conditional independence is violated, the tree model automatically performs partial pooling.}
\label{fig:tree_example}
\end{figure}

Following the default BART model specification, we model $f_0$ and $f_1$ as sums-of-trees,
\begin{align}
f_0(\bx, s) = \sum_{j = 1}^{m_0} g(\bx, s; \T_{0j}, \M_{0j}), 
\quad \text{and} \quad
f_1(\bx) = \sum_{j = 1}^{m_1} g(\bx; \T_{1j}, \M_{1j}),
\label{eq:tree_model}
\end{align}
where $\T_{0j}$ and $\T_{1j}$ are binary trees,  $\M_{0j}$ and $\M_{1j}$ are sets of parameters associated with the terminal nodes of the trees, and $g(\cdot)$ is a function that maps $(\bx, s)$ or $\bx$ to a parameter in $\M$ based on $\T$. This will be clear next.
We use $f_0$ as an example to explain the sum-of-trees model, and the model for $f_1$ follows the same logic, although information borrowing is not necessary for the treatment arm.
It is helpful to establish notation for a single tree model first. Let $\T_0$ denote a binary tree. Each interior node of $\T_0$ represents a decision rule, through which an $(\bx, s)$ pair is sent either left or right and eventually to a terminal node.
The terminal nodes define a partition of the $(\bx, s)$ space into subspaces.
Next, assume the $k$th terminal node is associated with a parameter $\mu_k$, which represents the mean response of the subgroup of observations that fall in that node.
Let $\M_0 = \{ \mu_{01}, \mu_{02}, \ldots, \mu_{0b} \}$ denote the set of such parameters, where $b$ is the number of terminal nodes. Given the tree model $(\T_0, \M_0)$ and a pair of observations $(\bx, s)$, $g(\bx, s; \T_0, \M_0)$ is defined as a single parameter value $\mu$ associated with the terminal node to which $(\bx, s)$ belongs.
The left panel in Figure \ref{fig:tree_example}(a) shows an example of a tree model $(\T_0, \M_0)$ with a one dimensional $x$ and a binary $s$, i.e., one covariate and one external data source ($s \in \{ 0, 1 \}$). The decision rules in the figure are the criteria for sending an $(x, s)$ pair to the left branch. In this example, we have $g(0.5, 0; \T_0, \M_0) = 0.694$.

The tree model adaptively pools information between the current trial and external data.  To see this, we consider two hypothetical trial examples with one covariate and a single external data source. In Example 1, we generate control outcomes (30 in the current trial, and 70 in the external data) as follows: $X \mid S = 0 \sim \N(0.7, 0.2^2)$, $X \mid S = 1 \sim \N(0.3, 0.4^2)$, and
$Y(0) \mid X \sim \N(1 -  X^2, 0.1^2)$.
Here, the distributions of $X$ in the current trial and external data are different.
This is typical, as a clinical trial usually has inclusion/exclusion criteria to enroll patients that are more likely to benefit from the new treatment, while the external data may contain observations from a wider population.
The control outcome is assumed to be conditionally independent of data source.
In Example 2, the only difference is that the control outcomes in the external data are generated from $Y(0) \mid X, S = 1 \sim \N(1.4 - 1.2 X^2, 0.1^2)$, indicating the control outcome and data source are not conditionally independent.
A simulated dataset under Example 1 is illustrated in Figure \ref{fig:tree_example}(a), as well as the fitted single tree model. The data source indicator $s$ does not appear in the decision rules, meaning the tree model is able to determine that the control outcome is conditionally independent of data source for all $x$ values.
All the $\mu$ values are estimated by pooling the trial and external data.
Figure \ref{fig:tree_example}(b) shows a simulated dataset under Example 2. The tree model in this case is able to identify some discrepancy across data sources. For the area of $x$ with the most different control outcomes and a sufficient amount of data, the $\mu$ values are estimated separately for the current trial and external data. For the areas of $x$ with similar control outcomes or less data, the $\mu$ values are still estimated by pooling the trial and external data. In summary, a regression tree model allows partial data pooling, adaptively improving the precision of the parameter estimates.

\begin{figure}[h!]
\begin{center}
\includegraphics[height=6.2cm]{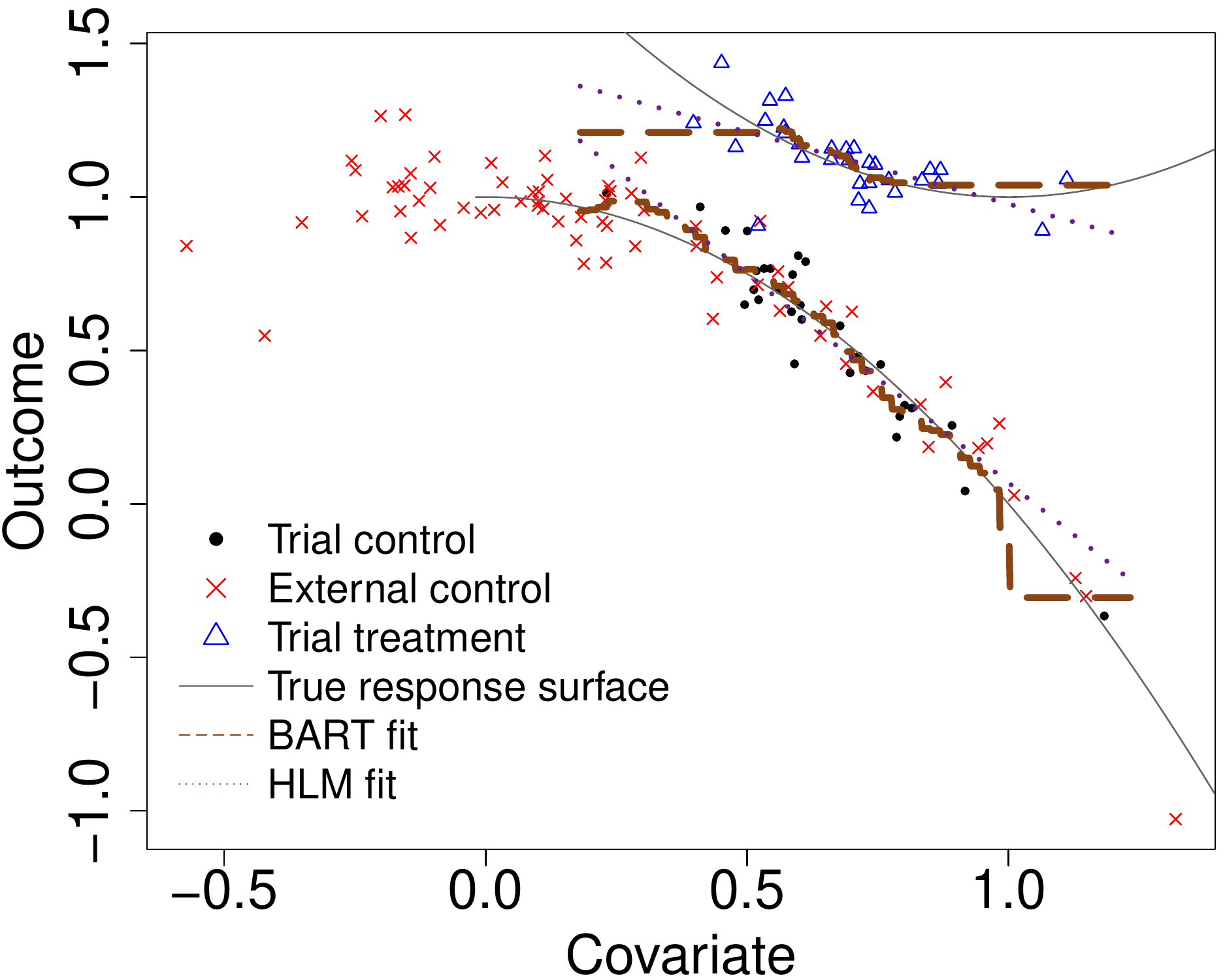}
\hspace{4mm}
\includegraphics[height=6.2cm]{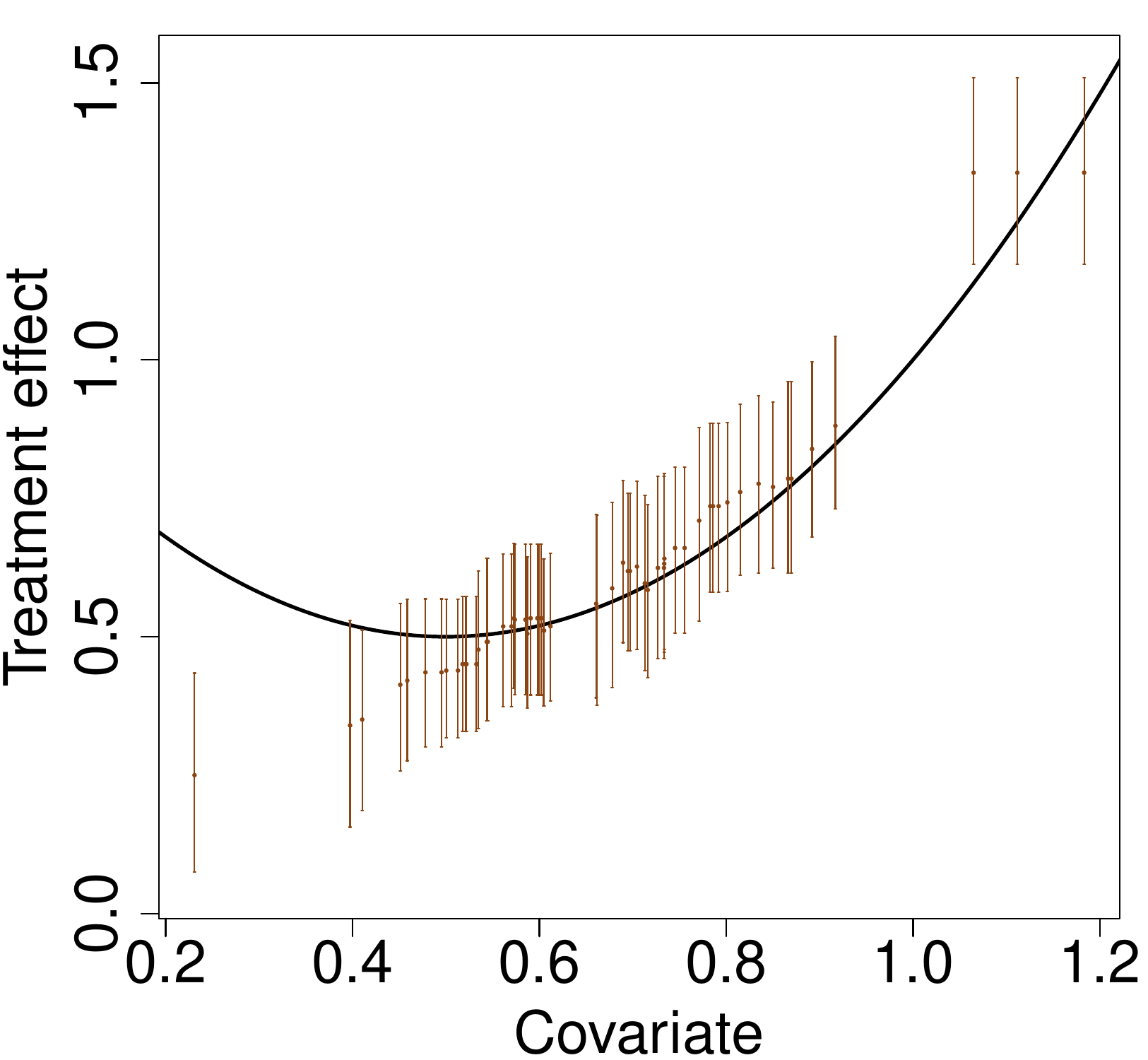}
\end{center}
\caption{Left panel: simulated trial and external data with BART and hierarchical linear model (HLM) fits.
Right panel: BART inference for the conditional treatment effects compared with the simulation truth.}
\label{fig:bart_example}
\end{figure}

Compared to a single tree model, a sum-of-trees model usually leads to better model fit and predictive capability.
This is why model \eqref{eq:tree_model} consists of $m_0$ and $m_1$ trees.
The idea is the following:  consider first fitting a single tree model to a dataset, denoted by $g(\bx, s; \T_{01}, \M_{01})$. The residuals can be calculated from $y - g(\bx, s; \T_{01}, \M_{01})$, and one can then fit the next tree model to these residuals. Repeating this procedure $m_0$ times, we get a total of $m_0$ trees. 
See \cite{chipman2010bart} for more details.  
Since each subtree allows partial information pooling, the sum-of-trees model also allows so.
Figure \ref{fig:bart_example} provides an illustration of a sum-of-trees model fit to the simulated dataset in Example 1, which nicely captures the covariate-outcome relationship.

The BART model also consists of a set of regularization priors over the tree parameters, which keeps the effect of each subtree small and further improves its performance. 
For example, for each binary tree $\T$, the probability that a node at depth $d~(= 0, 1, 2, \ldots)$ is nonterminal is given by $\rho (1+d)^{-\kappa}$. By default, $\rho = 0.95$ and $\kappa = 2$, which penalize deep and bushy trees.  The distributions on the splitting variable assignment and the splitting rule assignment conditional on the splitting variable are given uniform priors.
The numbers of subtrees, $m_0$ and $m_1$, are pre-specified with a default value of 200.
The $\mu$'s are given normal conjugate priors. Lastly, conjugate inverse-gamma distribution priors are placed on $\sigma_0^2$ and $\sigma_1^2$, the parameters of which are calibrated based on the residual standard deviations from least squares. Refer to \cite{chipman2010bart} for more details.

In \cite{chipman2010bart}, a Markov chain Monte Carlo algorithm is employed to implement posterior inference for a BART model. In particular, the algorithm generates $L$ draws from the joint posterior distribution of the model parameters, 
\begin{align*}
&\{ (\T_{01}^{(\ell)}, \M_{01}^{(\ell)}), \ldots, (\T_{0 m_0}^{(\ell)}, \M_{0 m_0}^{(\ell)}), \sigma_0^{(\ell)} \mid \ell = 1, 2, \ldots, L \} \quad \text{and} \\
&\{ (\T_{11}^{(\ell)}, \M_{11}^{(\ell)}), \ldots, (\T_{1 m_1}^{(\ell)}, \M_{1 m_1}^{(\ell)}), \sigma_1^{(\ell)} \mid \ell = 1, 2, \ldots, L \}.
\end{align*}
For a specific patient in the current trial ($s = 0$) with covariate values $\bx^*$, the posterior samples of $f_0(\bx^*, 0)$ and $f_1(\bx^*)$ can be respectively calculated based on
\begin{align*}
f_0^{(\ell)}(\bx^*, 0) = \sum_{j = 1}^{m_0} g(\bx, s; \T_{0j}^{(\ell)}, \M_{0j}^{(\ell)}), 
\quad \text{and} \quad
f_1^{(\ell)}(\bx^*) = \sum_{j = 1}^{m_1} g(\bx; \T_{1j}^{(\ell)}, \M_{1j}^{(\ell)}).
\end{align*}
The posterior samples of the conditional treatment effect at $\bx^*$, $\delta(\bx^*)$, are given by $\delta^{(\ell)}(\bx^*) = f_1^{(\ell)}(\bx^*) - f_0^{(\ell)}(\bx^*, 0)$ for $\ell = 1, 2, \ldots, L$ (see Equation \ref{eq:CATE3}). Figure \ref{fig:bart_example} (right panel) shows for a simulated dataset the posterior credible intervals of the conditional treatment effects (vertical line segments) evaluated at the $\bx^*$ values of the trial patients.

\paragraph{Estimation of the CATE}
Let $\{ \bx_{i}^* \mid i : S_i = 0 \}$ denote the set of covariates values for the patients in the current trial.
From Equation \eqref{eq:CATE2}, the posterior samples of $\Delta_{\C}$ can be calculated by
$\Delta_{\C}^{(\ell)} = \sum_{i: S_i = 0} \delta^{(\ell)}(\bx_i^*) / N_{\text{trial}}$.
The sample mean of $\{ \Delta_{\C}^{(\ell)}: \ell = 1, \ldots, L\}$ approximates the posterior mean of $\Delta_{\C}$, and the sample quantile of $\{ \Delta_{\C}^{(\ell)}: \ell = 1, \ldots, L\}$ approximates the quantile of the posterior distribution of $\Delta_{\C}$.

\paragraph{Estimation of the PATE}
If the goal is to estimate the PATE, the distribution of patient characteristics in the trial population, $[\bX \mid S = 0]$, also needs to be modeled. In this case, we propose to use the Bayesian bootstrap prior \citep{rubin1981bayesian}.
We assume that $\bX$ can only take the (discrete) values in $\{ \bx_{i}^* \mid i : S_i = 0 \}$. Further, let $\zeta_i$ be the probability associated with $\bx_i$, i.e., $\Pr(\bX = \bx_i^* \mid S = 0) = \zeta_i$, and $\sum \zeta_i = 1$.
With a Dirichlet distribution prior on the vector of $\zeta_i$'s, its posterior distribution is also a Dirichlet distribution.
By drawing $L$ posterior samples of the $\zeta_i$'s, the posterior samples of $\Delta_{\P}$ can be calculated by
$\Delta_{\P}^{(\ell)} = \sum_{i: S_i = 0} \left[ \zeta_i^{(\ell)} \delta^{(\ell)}(\bx_i^*) \right]$.

\section{Simulation Studies}
\label{sec:simulation}

We conduct several simulation studies to evaluate the operating characteristics of the proposed BART method.
We focus on the estimation of the CATE, because the true PATE can be hard to compute in general as it can involve complicated multiple integral (see Equation \ref{eq:CATE2}), making it challenging to check how the estimated PATE deviates from the truth.
In Section \ref{sec:sim_setup}, we consider scenarios with a single external data source ($J = 1$) under the assumption of conditional independence (of the control outcome and data source).
In practice, it is common that only a single external data source is available for borrowing.
In Section \ref{sec:non_independent}, we explore scenarios in which conditional independence is violated. 
In Section \ref{sec:multiple_external}, we examine the performance of the proposed method with multiple external data sources ($J > 1$).
In all simulation scenarios, we assume that the individuals in the current study are randomized to treatment and control, and those in the external data are all treated by the control drug.

\subsection{Simulation Setup}
\label{sec:sim_setup}

We first consider the following three simulation scenarios, under which the true data-generating models satisfy conditional independence. 
For each scenario, assume $50$ individuals are enrolled in the current study, and data of 200 individuals from a single external data source ($J = 1$) are available.

\paragraph{Scenario 1.}
We consider an illustrative example with an one-dimensional covariate $X$.
We assume the distributions of $X$ in the current trial and external data are different, which is common in practice.
Specifically, we generate $X$ as follows,
\begin{align*}
X \mid S = 0 \sim \N(0.7, 0.2^2), \quad \text{and} \quad
X \mid S = 1 \sim \N(0.3, 0.4^2).
\end{align*}
The treatment assignment in the current trial is random, 
\begin{align*}
T \mid S = 0 \sim \Ber(0.5).
\end{align*}
Lastly, we assume the outcome has a non-linear relationship with $X$, generated from
\begin{align*}
Y \mid T, X \sim
\N\left(1  - 0.16 T + T  (X - 1)^2  -  (1-T)  X^2, \, 0.1^2 \right),
\end{align*}
The true CATE depends on the generated covariate values, and the true PATE is $\E[\E[Y(1) -  Y(0) \mid \bX, S = 0]] = 0.5$.
Illustrations of the simulated data can be found in Figures \ref{fig:tree_example} and \ref{fig:bart_example}.

\paragraph{Scenario 2.}

We consider a scenario with $Q = 4$ covariates, which are generated by first drawing
\begin{align*}
(X_1, X_2, X_3, \tilde{X}_4) \mid S = 0 &\sim \N_4 (\bmu_{X0}, \sigma_{X0}^2 \bbOmega_{X}),  \\
(X_1, X_2, X_3, \tilde{X}_4) \mid S = 1 &\sim \N_4 (\bmu_{X1}, \sigma_{X1}^2 \bbOmega_{X}),
\end{align*}
and then setting
$X_4 \sim \Ber \left[ \Phi(\tilde{X}_4 - 0.5) \right]$ for both groups.
In this way, we obtain a combination of continuous and binary covariates, similar to the real-data application.
Here, $\N_Q( \cdot,\cdot)$ denotes a $Q$-variate normal distribution, $\bmu_{X0} = (0.7, \ldots, 0.7)$, $\sigma_{X0} = 0.2$, $\bmu_{X1} = (0.3, \ldots, 0.3)$, $\sigma_{X1} = 0.4$,
and $\bbOmega_{X}$ is a correlation matrix with off-diagonal elements randomly sampled from $(0.1, 0.4, 0.7, -0.3)$ with probabilities $(0.4, 0.3, 0.1, 0.2)$.
The distributions of $\bX$ in the current trial and external data have some overlap but are not identical.

The treatment and control outcomes are generated respectively as follows,
\begin{align*}
Y \mid T = 1, \bX &\sim
\N \left( {\bX}^{\top} \bbeta_{1} + 5 , \, 0.5^2 \right), \\
Y \mid T = 0,  \bX &\sim
\N \left( \exp(\bX^{\top} \bbeta_{0}), \, 0.5^2 \right),
\end{align*}
assuming that the control outcome has a nonlinear relationship with $\bX$.
The  coefficients in $\bbeta_1$ and $\bbeta_0$ are randomly sampled from $(0.1, 0.7)$ with probabilities $(0.3, 0.7)$.
The true CATE depends on the generated covariate values, and the true PATE in this case is hard to compute as it involves complicated multiple integral.

\paragraph{Scenario 3.}

We consider a scenario where the treatment and control outcomes have no relationship with the covariates.
The goal of this simulation is to assess the loss of efficiency from using an unnecessarily complex modeling approach.
Again, consider $Q = 4$ covariates.
For both the current trial and external data, the covariates are generated by first drawing
\begin{align*}
(X_1, X_2, X_3, \tilde{X}_4) \sim \N_4(\bmu_{X}, \sigma_{X}^2 \bbOmega_X), 
\end{align*}
where $\bmu_{X} = (0.5, \ldots, 0.5)$, $\sigma_{X} = 0.5$, and $\bbOmega_X$ is generated in the same way as in Scenario 2.
Then, we sample
$X_4 \sim \Ber \left[ \Phi(2 \tilde{X}_4 - 1) \right]$,  where $\Phi$ denotes the cumulative distribution function of the standard normal distribution.
Finally, the outcomes are generated from 
\begin{align*}
Y \mid T \sim \N\left(0.2 + 0.5 T , \, 0.1^2 \right). 
\end{align*}
In this case, the true CATE and true PATE are both 0.5 and do not depend on the empirical distribution of $\bX$.

\subsection{Competing Methods and Performance Metrics}

We compare the proposed BART method with the following alternatives:
\begin{enumerate}[noitemsep,nolistsep,leftmargin=.5in]
\item (HLM) A hierarchical linear model for the response surface of the following form,
\begin{align*}
\begin{split}
[Y \mid T = 0, \bX = \bx, S = s] &\sim \N(\alpha_{0s} + \bx^{\top} \bbeta_{0s}, \sigma_{0}^2), \\
[Y \mid T = 1, \bX = \bx, S = 0] &\sim \N(\alpha_1 + \bx^{\top} \bbeta_{1}, \sigma_{1}^2).
\end{split}
\end{align*}
The regression coefficients are given priors 
\begin{align*}
&\alpha_{0s} \sim \N(\tilde{\alpha}_0, \tau_{\alpha}^2), \quad \bbeta_{0s} \sim \N_Q\left[\tilde{\bbeta}_0, \text{diag}(\tau_{\beta_1}^2, \ldots, \tau_{\beta_Q}^2)\right], \quad \sigma_0^2 \sim \IG(\nu, \nu), \\
&\tilde{\alpha}_0 \sim \N(0, 10^2), \quad \tilde{\bbeta}_0 \sim \N_Q(\bm 0, 10^2 \bbI), \quad  \tau_{\alpha}^2, \tau_{\beta_1}^2, \ldots, \tau_{\beta_Q}^2 \sim \IG(\nu, \nu), \\
&\alpha_1 \sim \N(0, 10^2), \quad \bbeta_1 \sim \N_Q(\bm 0, 10^2 \bbI), \quad \sigma_1^2 \sim \IG(\nu, \nu),
\end{align*}
where $\nu = 10^{-4}$.
Posterior inference for the CATE follows a similar procedure as in Section \ref{sec:model}.
\item (NNHM) A normal-normal hierarchical model that does not take into account covariates,
\begin{align*}
\begin{split}
[Y \mid T = 0, S = s] &\sim \N( \alpha_{0s}, \sigma_{0}^2), \\
[Y \mid T = 1, S = 0] &\sim \N(\alpha_{1}, \sigma_{1}^2).
\end{split}
\end{align*}
The model parameters are given priors 
\begin{align*}
&\alpha_{0s} \sim \N(\tilde{\alpha}_0, \tau_{\alpha}^2), \quad \tilde{\alpha}_0 \sim \N(0, 10^2),  \quad \tau_{\alpha}^2 \sim \IG(\nu, \nu), \\
&\sigma_0^2 \sim \IG(\nu, \nu), \quad \alpha_1 \sim \N(0, 10^2),  \quad \sigma_1^2 \sim \IG(\nu, \nu).
\end{align*}
Inference for the CATE is based on the posterior distribution of $(\alpha_1 - \alpha_{00})$.
\item For each of the BART, HLM, and NNHM methods, we also consider a version that does not make use of external data. These are denoted by BART$-$, HLM$-$, and NNHM$-$, respectively.  For example, BART$-$ models
\begin{align*}
\begin{split}
[Y \mid T = 0, \bX = \bx, S = 0] &\sim \N\left(f_0(\bx), \sigma_{0}^2\right), \\
[Y \mid T = 1, \bX = \bx, S = 0] &\sim \N\left(f_1(\bx), \sigma_{1}^2\right),
\end{split}
\end{align*}
where $f_0$ and $f_1$ are sums-of-trees similar to Equation \eqref{eq:tree_model}.  Therefore, external data (those with $S \geq 1$) are ignored in this case. 
\end{enumerate}
The performance of each method is evaluated based on the following metrics:
\begin{enumerate}[noitemsep,nolistsep,leftmargin=.5in]
\item Bias and root mean squared error (RMSE) in estimating the CATE, given by
\begin{align*}
\text{Bias} = \hat{\Delta}_{\C} - \Delta_{\C}^{\true} 
\quad \text{and} \quad
\text{RMSE} = \sqrt{\frac{1}{L} \sum_{\ell = 1}^L \left( \Delta_{\C}^{(\ell)} - \Delta_{\C}^{\true} \right)^2},
\end{align*}
respectively, where $\hat{\Delta}_{\C}$ is the posterior mean of $\Delta_{\C}$.
\item Coverage and length of the 95\% credible interval (CI) of $\Delta_{\C}$.
\item Precision in estimation of heterogeneous effects (PEHE, \citealp{hill2011bayesian}), operationalized as 
\begin{align*}
\text{PEHE} = \sqrt{\frac{1}{N_{\text{trial}}} \sum_{i: S_i = 0} [ \hat{\delta}(\bx_{i}^*) - \delta^{\true}(\bx_{i}^*) ]^2 }, 
\end{align*}
where $\hat{\delta}(\bx_{i}^*)$ is the posterior mean of $\delta(\bx_{i}^*)$.
A smaller PEHE indicates a better capture of heterogeneous treatment effects.
\item Results of the hypothesis tests:
\begin{alignat*}{4}
&\text{Test 1:} \quad &&H_{10}: \Delta_{\C} \geq \Delta_{\C}^{\true}  \quad &&\text{vs.} \quad &&H_{11}: \Delta_{\C} < \Delta_{\C}^{\true};  \\
&\text{Test 2:} \quad &&H_{20}: \Delta_{\C} \geq \Delta_{\C}^{\true} - d \quad &&\text{vs.} \quad  &&H_{21}: \Delta_{\C} < \Delta_{\C}^{\true} - d.
\end{alignat*}
Here, $H_{10}$ is rejected if $\Pr(\Delta_{\C} > \Delta_{\C}^{\true} \mid \text{data}) > 0.95$, which is a false positive (type I error); $H_{20}$ is rejected if $\Pr(\Delta_{\C} > \Delta_{\C}^{\true} - d \mid \text{data}) > 0.95$, which is a true positive (for some $d > 0$). The fraction of times $H_{10}$ is rejected reflects the type I error rate of a method, while that for $H_{20}$ reflects the power.
We set $d = 0.08$ for Scenarios 1 and 3, and $d = 0.25$ for Scenario 2. These values are chosen to facilitate the comparison of the powers for different methods.
\end{enumerate}

\subsection{Simulation Results}

For each simulation scenario, we generate 500 datasets and fit the six models to each dataset. For each method, we run MCMC simulation for 1,100 iterations with the first 100 draws discarded as burn-in.

Table \ref{tbl:sim_tbl1} summarizes the simulation results. The values are averages over repeated simulations. 
The CATE slightly differs for each simulated dataset, because each time a distinct set of covariate values are generated. The average CATE over repeated simulations is also reported in Table \ref{tbl:sim_tbl1}.
The average CATE, bias, RMSE, CI length and PEHE have been multiplied by 100 to facilitate comparison between methods.
Overall, the BART method leads to the lowest RMSE, shortest CI length, lowest PEHE, and highest power among all methods across scenarios. The CI coverage and type I error rate of BART do not give rise to concerns. An exact match with the nominal frequentist coverage probability is not expected as all methods considered are Bayesian.
Since the current trial is randomized and controlled, methods ignoring external data (BART$-$, HLM$-$ and NNHM$-$) can estimate the CATE with nearly no bias. By adjusting for prognostic and predictive covariates, the BART$-$ and HLM$-$ methods have better powers in Scenarios 1 and 2 compared to NNHM$-$.
Furthermore, since the control outcome is conditionally independent of data source, borrowing from external control data should ideally improve the estimation of the CATE. However, we can see this is only the case for BART but not for HLM and NNHM. First, note that in Scenarios 1 and 2, the control outcome is not \emph{marginally independent} of data source when not conditioning on the covariates. Therefore, the NNHM method, which does not adjust for covariates, has higher bias and potentially lower power in Scenarios 1 and 2 compared to NNHM$-$. Second, in Scenarios 1 and 2, the relationships between the covariates and control outcomes are nonlinear. Therefore, the HLM method, which shrinks the slope and intercept coefficients, also does not compare favorably to HLM$-$ in these scenarios. Finally, the BART method allows flexible information pooling and performs better than all the other methods. Even when the control outcomes have no relationship with the covariates (Scenario 3), the BART method does not lead to a loss of efficiency.

\begin{table}[h!]
\begin{center}
\begin{tabular}{lrrrrrrr}
\toprule
\multicolumn{1}{c}{Model}  & \multicolumn{1}{c}{Bias} & \multicolumn{1}{c}{RMSE} & \multicolumn{1}{c}{\% Cover} & \multicolumn{1}{c}{CI length} & \multicolumn{1}{c}{PEHE} & \multicolumn{1}{c}{\% Rej. 1} & \multicolumn{1}{c}{\% Rej. 2} \\ \midrule
\multicolumn{8}{c}{Scenario 1 (Avg. CATE = 50.18)} \\
BART & $-0.48$ & $4.27$ & $95.6$ & $13.00$ & $8.15$ & $2.0$ & $77.4$ \\
HLM & $0.19$ & $6.00$ & $100.0$ & $20.73$ & $13.21$ & $0.2$ & $46.0$ \\
NNHM & $-2.54$ & $10.50$ & $96.6$ & $31.43$ & $20.40$ & $0.8$ & $14.2$ \\
BART$-$ & $-0.52$ & $4.81$ & $93.8$ & $14.10$ & $11.79$ & $3.4$ & $71.6$ \\
HLM$-$ & $-0.45$ & $4.29$ & $96.0$ & $13.31$ & $11.04$ & $1.8$ & $75.8$ \\
NNHM$-$ & $-0.21$ & $9.29$ & $96.0$ & $28.29$ & $20.08$ & $2.6$ & $26.0$ \\
\midrule
\multicolumn{8}{c}{Scenario 2 (Avg. CATE = 153.34)} \\
BART & $0.59$ & $20.56$ & $93.8$ & $56.15$ & $35.20$ & $6.2$ & $55.8$ \\
HLM & $-2.45$ & $22.50$ & $95.4$ & $65.27$ & $45.96$ & $3.4$ & $40.2$ \\
NNHM & $4.71$ & $40.55$ & $94.2$ & $115.36$ & $155.10$ & $7.0$ & $27.2$ \\
BART$-$ & $-0.89$ & $22.41$ & $93.6$ & $61.29$ & $49.02$ & $5.6$ & $48.8$ \\
HLM$-$ & $-1.22$ & $22.32$ & $95.8$ & $65.78$ & $40.58$ & $3.4$ & $43.6$ \\
NNHM$-$ & $1.37$ & $47.23$ & $98.6$ & $150.74$ & $154.94$ & $1.8$ & $11.0$ \\
\midrule
\multicolumn{8}{c}{Scenario 3 (Avg. CATE = 50.00)} \\
BART & $0.10$ & $3.75$ & $94.6$ & $10.63$ & $4.37$ & $5.6$ & $90.6$ \\
HLM & $0.09$ & $3.99$ & $97.0$ & $11.85$ & $5.59$ & $3.2$ & $87.0$ \\
NNHM & $0.08$ & $3.68$ & $96.4$ & $10.97$ & $2.10$ & $3.4$ & $90.6$ \\
BART$-$ & $0.09$ & $3.84$ & $93.4$ & $10.68$ & $4.83$ & $6.4$ & $90.2$ \\
HLM$-$ & $0.10$ & $4.22$ & $96.4$ & $12.61$ & $6.51$ & $3.4$ & $84.4$ \\
NNHM$-$ & $0.10$ & $3.98$ & $97.0$ & $11.77$ & $2.31$ & $3.8$ & $87.6$ \\
\bottomrule
\end{tabular}
\end{center}
\caption{Simulation results under the three simulation scenarios with a single external data source under the assumption of conditional independence. Values shown are averages over 500 repeated simulations. The average CATE, bias, RMSE, CI length and PEHE have been multiplied by 100 to facilitate comparison between methods. \% Rej. 1 and \% Rej. 2 represent the percentages of times $H_{10}$ and $H_{20}$ are rejected, respectively.} 
\label{tbl:sim_tbl1}
\end{table}

\subsection{Violation of Conditional Independence}
\label{sec:non_independent}

We conduct additional simulation studies to assess the performance of the proposed BART method when the control outcome is not conditionally independent of data source. Specifically, in Scenario 1, we now generate the external control outcome from
\begin{align*}
Y \mid T = 0, X, S = 1 \sim \N\left(1.4   - 1.2  X^2, \, 0.1^2 \right).
\end{align*}
Similarly, in Scenario 2, we generate
\begin{align*}
Y \mid T = 0, \bX, S = 1 \sim \N \left( \exp[\bX^{\top} (\bbeta_{0} + \bbeta_{0}^\text{diff}) ], \, 0.5^2 \right),
\end{align*}
where the  coefficients in $\bbeta_{0}^\text{diff}$ are randomly sampled from $(0.2, -0.2, 0)$ with probabilities $(0.3, 0.3, 0.4)$, conditioning on $\bbeta_{0}^\text{diff} \neq \bm 0$.
In Scenario 3,
\begin{align*}
Y \mid T = 0, S = 1 \sim \N\left(0.4, \, 0.1^2 \right).
\end{align*}
The other simulation settings, including the simulated trial data, are kept unchanged. 
To quantify the degree of heterogeneity between current and external control data, we calculate the CATE discrepancy by 
\begin{align*}
\frac{1}{N_{\text{trial}}} \sum_{i: S_i = 0} \left( \E[Y \mid T = 0, \bX_i, S = 0] - \E[Y \mid T = 0, \bX_i, S = 1] \right).
\end{align*}
A positive (or negative) CATE discrepancy means that the CATE calculated based on the response surface of the external control data is greater (or smaller) than that of the current control data.
Therefore, borrowing in the presence of a positive (or negative) CATE discrepancy would likely result in an overestimate (or underestimate) of the CATE.

The average CATE discrepancy over repeated simulations under each scenario is  reported in Table \ref{tbl:sim_tbl2}.
Since the current trial data remain unchanged, BART$-$, HLM$-$ and NNHM$-$ yield identical inference thus are not included in Table \ref{tbl:sim_tbl2}.
In Scenarios 1 and 3, the average CATE discrepancy is negative. Therefore, we can observe a reduction in power for every method compared to Table \ref{tbl:sim_tbl1}. Interestingly, under Scenario 1, BART still has better power compared to BART$-$ even in the non-independent case (with  a negative CATE discrepancy). This is because BART allows partial information pooling. For areas of the covariate space in which the trial and external data are relatively homogeneous (in the case of Scenario 1, for $x$ around $\sqrt{2}$), BART still pools the data to produce more precise estimates.
In Scenario 2, the average CATE discrepancy is positive, leading to some inflation of the type I error rate. However, since BART borrows information in a judicious manner, the amount of type I error rate inflation is not severe.

\begin{table}[h!]
\begin{center}
\begin{tabular}{lrrrrrrr}
\toprule
\multicolumn{1}{c}{Model}  & \multicolumn{1}{c}{Bias} & \multicolumn{1}{c}{RMSE} & \multicolumn{1}{c}{\% Cover} & \multicolumn{1}{c}{CI length} & \multicolumn{1}{c}{PEHE} & \multicolumn{1}{c}{\% Rej. 1} & \multicolumn{1}{c}{\% Rej. 2} \\ \midrule
\multicolumn{8}{c}{Scenario 1 (Avg. CATE = 50.10, CATE discrepancy = $-29.35$)} \\
BART & $-0.47$ & $4.38$ & $96.4$ & $13.45$ & $8.62$ & $2.2$ & $74.8$ \\
HLM & $-0.52$ & $6.65$ & $100.0$ & $23.49$ & $11.72$ & $0.0$ & $23.8$ \\
NNHM & $-1.34$ & $10.75$ & $99.2$ & $34.57$ & $20.15$ & $0.4$ & $12.4$ \\
\midrule
\multicolumn{8}{c}{Scenario 2 (Avg. CATE = 152.48, CATE discrepancy = 113.91)} \\
BART & $1.28$ & $20.91$ & $90.8$ & $55.60$ & $40.35$ & $9.0$ & $58.0$ \\
HLM & $-1.63$ & $21.91$ & $95.0$ & $62.15$ & $45.95$ & $5.2$ & $44.0$ \\
NNHM & $2.66$ & $35.74$ & $87.4$ & $92.35$ & $154.99$ & $10.6$ & $37.6$ \\
\midrule
\multicolumn{8}{c}{Scenario 3 (Avg. CATE = 50.00, CATE discrepancy = $-20.00$)} \\
BART & $-0.61$ & $3.83$ & $93.6$ & $10.69$ & $4.58$ & $2.8$ & $84.8$ \\
HLM & $-0.16$ & $4.14$ & $95.8$ & $12.17$ & $5.78$ & $2.8$ & $82.0$ \\
NNHM & $-0.11$ & $3.95$ & $96.2$ & $11.56$ & $2.32$ & $3.8$ & $86.4$ \\
\bottomrule
\end{tabular}
\end{center}
\caption{Simulation results under the three simulation scenarios with a single external data source, assuming violation of conditional independence. Values shown are averages over 500 repeated simulations. The average CATE, bias, RMSE, CI length and PEHE have been multiplied by 100 to facilitate comparison between methods. \% Rej. 1 and \% Rej. 2 represent the percentages of times $H_{10}$ and $H_{20}$ are rejected, respectively.} 
\label{tbl:sim_tbl2}
\end{table}

\subsection{Multiple External Data Sources}
\label{sec:multiple_external}

We examine the performances of the methods with multiple external data sources ($J = 4$). 
Assume each external data source has data of 50 individuals.  The external control data are generated as follows. For Scenario 1, 
\begin{align*}
&X \mid S \in \{ 1, 3 \} \sim \N(0.7, 0.2^2), \quad
X \mid S \in \{ 2, 4 \} \sim \N(0.3, 0.4^2), \\
&Y \mid T = 0, X, S \in \{ 1, 2 \} \sim
\N\left(1 - X^2, \, 0.1^2 \right),  \; \text{and}\\
&Y \mid T = 0, X, S \in \{ 3, 4 \} \sim
\N\left(1.4 - 1.2 X^2, \, 0.1^2 \right).
\end{align*}
For Scenario 2,
\begin{align*}
&(X_1, X_2, X_3, \tilde{X}_4) \mid S \in \{ 1, 3 \} \sim \N_4 (\bmu_{X0}, \sigma_{X0}^2 \bbOmega_{X}), \\
&(X_1, X_2, X_3, \tilde{X}_4) \mid S \in \{ 2, 4 \} \sim \N_4 (\bmu_{X1}, \sigma_{X1}^2 \bbOmega_{X}), \\
&Y \mid T = 0, \bX, S \in \{ 1, 2 \} \sim \N \left( \exp(\bX^{\top} \bbeta_{0}), \, 0.5^2 \right),  \; \text{and}\\
&Y \mid T = 0, \bX, S \in \{ 3, 4 \} \sim \N \left( \exp[\bX^{\top} (\bbeta_{0} + \bbeta_{0}^\text{diff}) ], \, 0.5^2 \right).
\end{align*}
Lastly, for scenario 3,
\begin{align*}
&(X_1, X_2, X_3, \tilde{X}_4) \sim \N_4(\bmu_{X}, \sigma_{X}^2 \bbOmega_X), \\
&Y \mid T = 0, S \in \{ 1, 2 \} \sim
\N\left(0.2, \, 0.1^2 \right),  \; \text{and} \\
&Y \mid T = 0, S \in \{ 3, 4 \} \sim
\N\left(0.4, \, 0.1^2 \right).
\end{align*}
The other simulation settings, including the simulated trial data, are kept unchanged. 
From the data-generating process, we can see that 
$[Y \mid T = 0, \bX, S = 0 ] \stackrel{d}{=} [Y \mid T = 0, \bX, S = s]$ holds for $s \in \{1, 2\}$ but not for $s \in \{3, 4\}$.
In this way, we get a mixture of control data that are conditionally independent of data sources and those that are not.

All the BART, HLM and NNHM methods can readily accommodate multiple external data sources. Table \ref{tbl:sim_tbl3} summarizes the simulation results.
The performances of the methods remain similar to those under the previous settings, with BART outperforming the others.

\begin{table}[h!]
\begin{center}
\begin{tabular}{lrrrrrrr}
\toprule
\multicolumn{1}{c}{Model}  & \multicolumn{1}{c}{Bias} & \multicolumn{1}{c}{RMSE} & \multicolumn{1}{c}{\% Cover} & \multicolumn{1}{c}{CI length} & \multicolumn{1}{c}{PEHE} & \multicolumn{1}{c}{\% Rej. 1} & \multicolumn{1}{c}{\% Rej. 2} \\ \midrule
\multicolumn{8}{c}{Scenario 1 (Trial CATE = 50.10)} \\
BART & $-0.28$ & $4.32$ & $94.2$ & $12.63$ & $8.47$ & $5.2$ & $76.8$ \\
HLM & $-0.27$ & $5.39$ & $99.8$ & $17.99$ & $11.08$ & $0.6$ & $51.2$ \\
NNHM & $-1.71$ & $9.96$ & $98.0$ & $30.38$ & $20.18$ & $2.0$ & $15.4$ \\
\midrule
\multicolumn{8}{c}{Scenario 2 (Trial CATE = 152.48)} \\
BART & $0.79$ & $20.37$ & $92.4$ & $56.67$ & $36.24$ & $7.0$ & $56.0$ \\
HLM & $-0.91$ & $21.41$ & $95.0$ & $62.98$ & $45.09$ & $3.4$ & $43.4$ \\
NNHM & $5.95$ & $39.56$ & $95.0$ & $112.96$ & $154.42$ & $7.6$ & $27.6$ \\
\midrule
\multicolumn{8}{c}{Scenario 3 (Trial CATE = 50.00)} \\
BART & $-0.34$ & $3.83$ & $92.0$ & $10.54$ & $4.35$ & $5.4$ & $85.6$ \\
HLM & $-0.18$ & $4.15$ & $92.6$ & $11.86$ & $5.30$ & $3.8$ & $80.8$ \\
NNHM & $-0.17$ & $3.97$ & $93.8$ & $11.39$ & $2.36$ & $4.6$ & $84.0$ \\
\bottomrule
\end{tabular}
\end{center}
\caption{Simulation results under the three simulation scenarios with four external data sources. Conditional independence holds for only two of these data sources. Values shown are averages over 500 repeated simulations. The average CATE, bias, RMSE, CI length and PEHE have been multiplied by 100 to facilitate comparison between methods. \% Rej. 1 and \% Rej. 2 represent the percentages of times $H_{10}$ and $H_{20}$ are rejected, respectively.} 
\label{tbl:sim_tbl3}
\end{table}

\section{Application to an Acupuncture Trial}
\label{sec:real_data}

We illustrate the practical application of the proposed method based on the randomized controlled trial of acupuncture in \cite{vickers2004acupuncture}. 
The trial used randomized minimization to allocate 401 headache patients  to an acupuncture group (205 patients) and a control group (standard care, 196 patients).  
Patients randomized to acupuncture received acupuncture treatments over three months in addition to standard care.
Several outcome measures, such as the headache score on a 0--100 scale, were assessed at baseline, three, and 12 months.  A total of 161 patients in the acupuncture arm and 140 patients in the control arm completed the 12-month followup.
The available baseline characteristics of the patients were age, sex, headache type (tension or migraine), and number of years of headache disorder (chronicity).
The patient-level data of this trial were made available by \cite{vickers2006whose}.

\begin{figure}[h!]
\begin{center}
\begin{subfigure}{.32\textwidth}
\includegraphics[width=\textwidth]{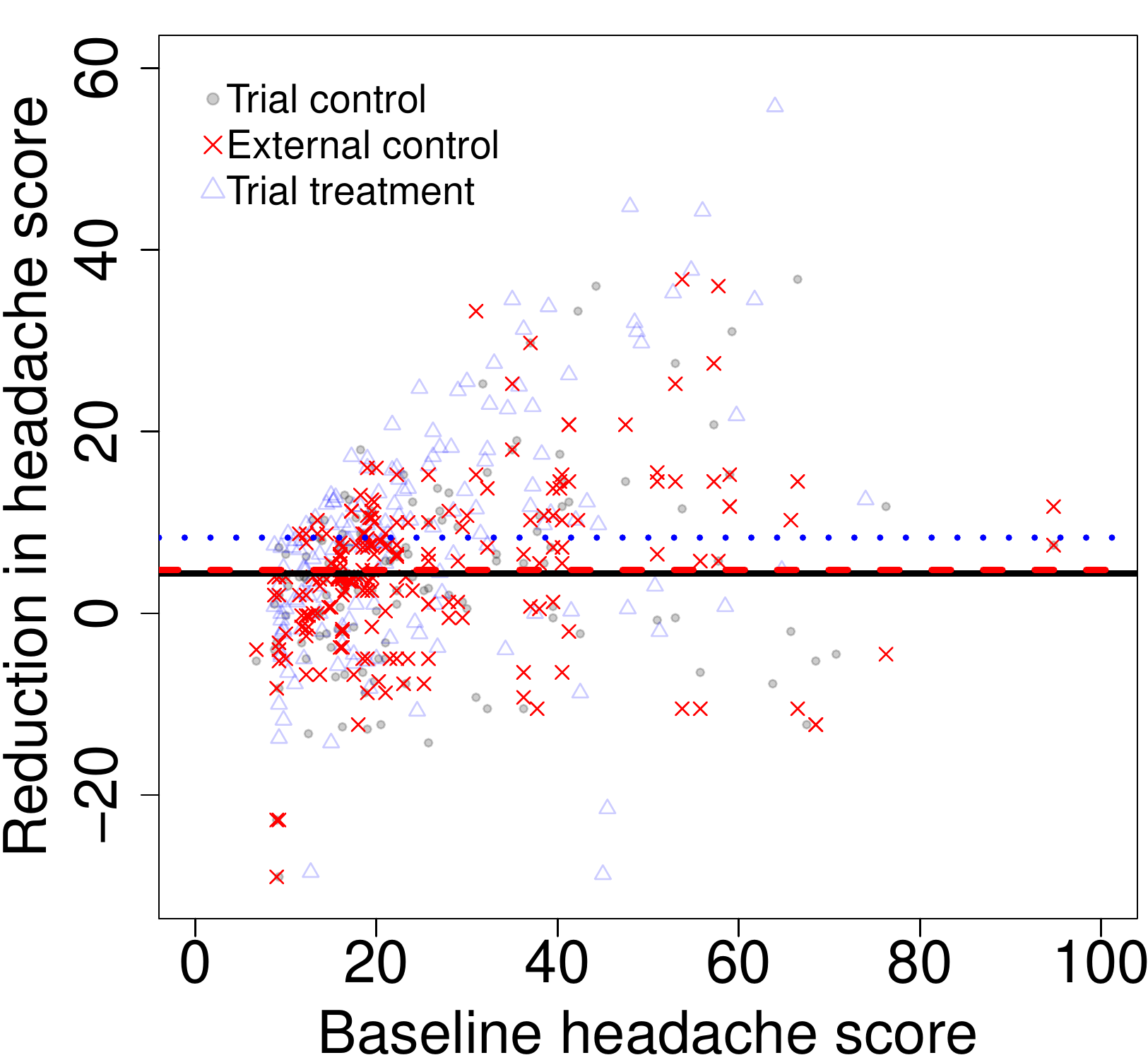}
\caption{Scenario 1}
\end{subfigure}
\begin{subfigure}{.32\textwidth}
\includegraphics[width=\textwidth]{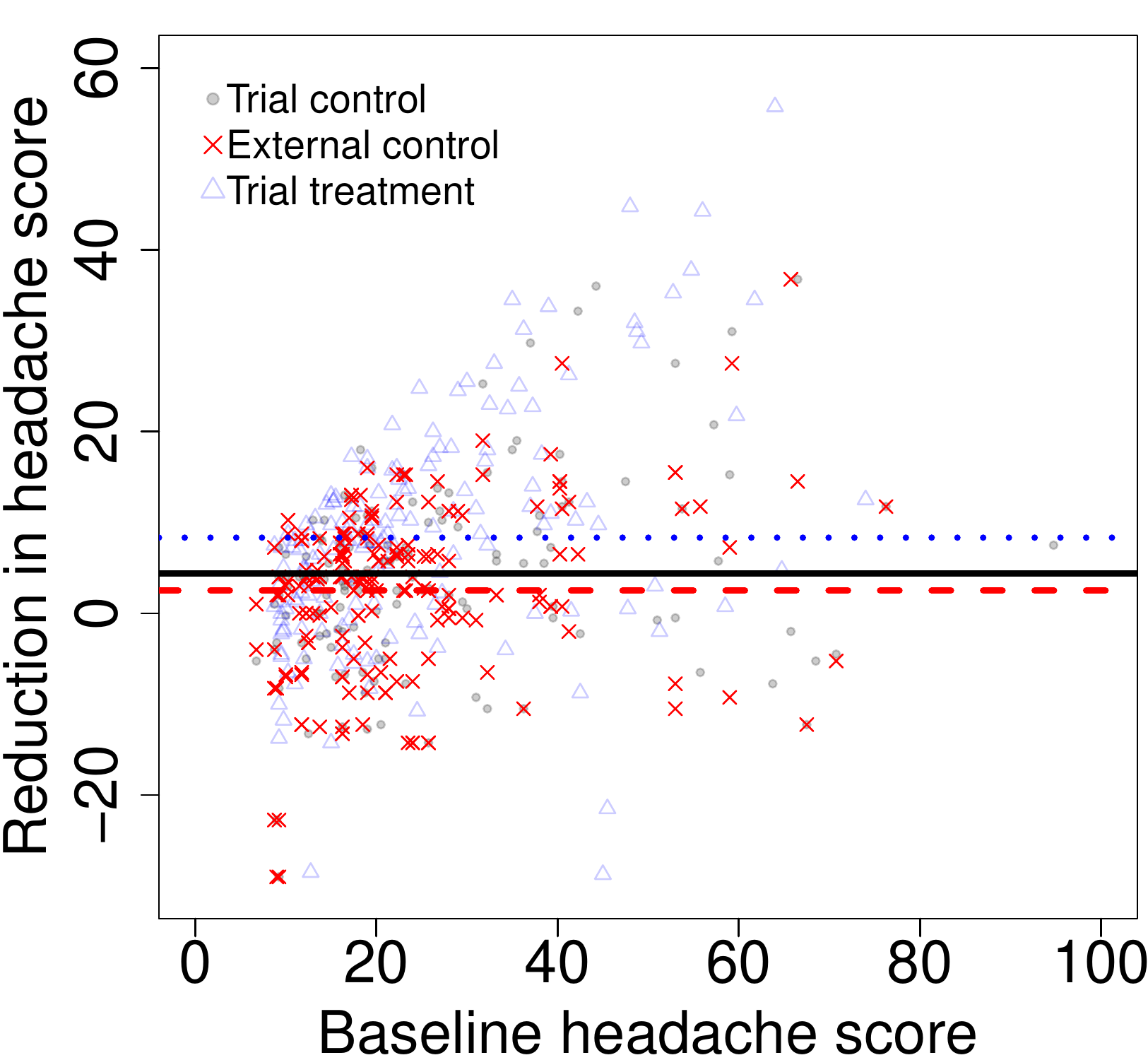}
\caption{Scenario 2}
\end{subfigure}
\begin{subfigure}{.32\textwidth}
\includegraphics[width=\textwidth]{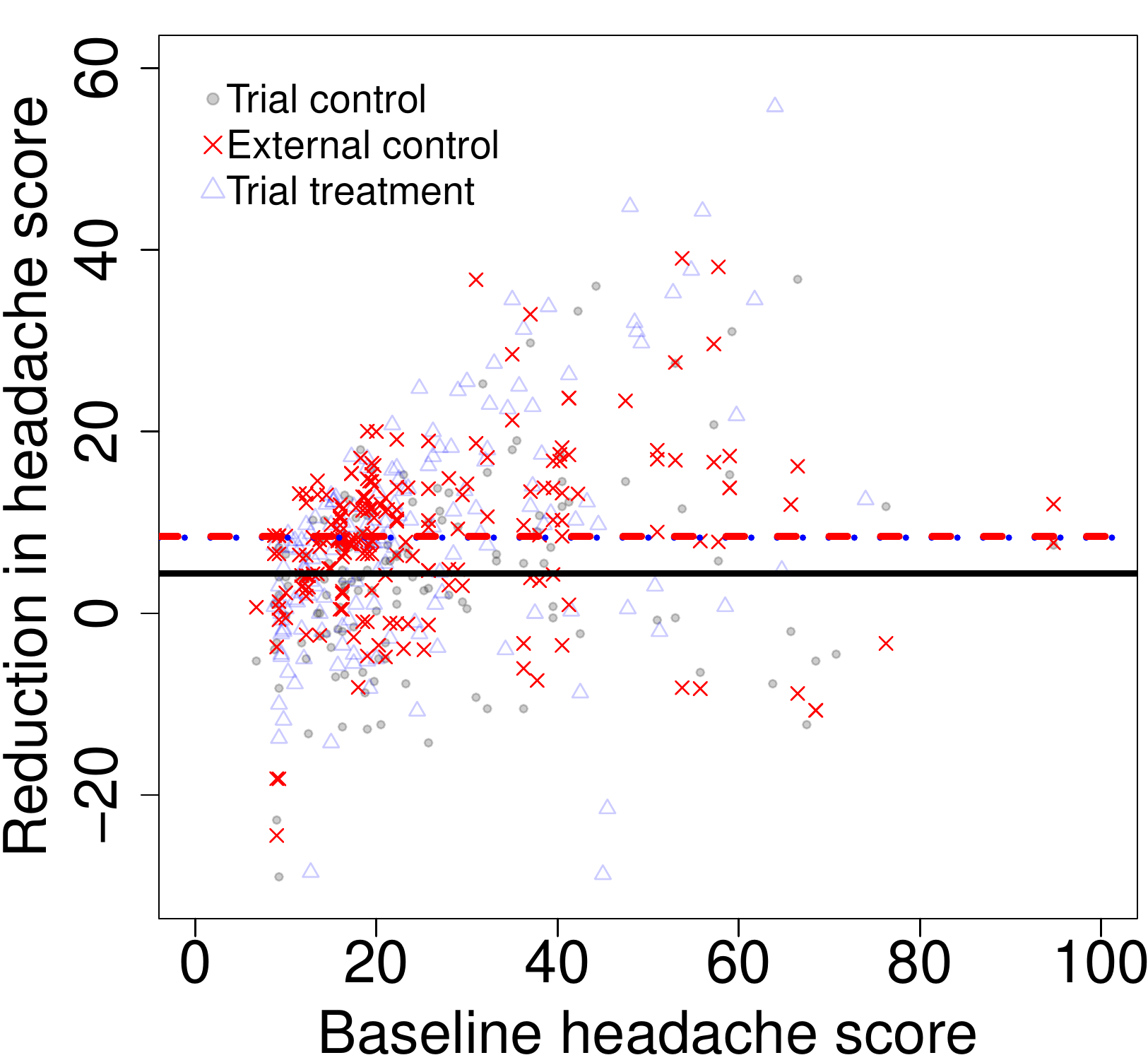}
\caption{Scenario 3}
\end{subfigure}
\end{center}
\caption{Trial data and simulated external control data under each scenario. The solid, dashed, and dotted horizontal lines represent the overall average reductions in headache scores in the trial control, external control, and trial treatment groups, respectively. }
\label{fig:real_data}
\end{figure}

For the purpose of our illustration, we consider the decrease in the headache score at 12 months compared to that at baseline as the primary outcome ($Y$).  The covariates ($\bX$) are baseline headache score, age, sex, headache type, and chronicity, where sex and headache type are binary covariates, and the others are continuous.
Also, since our focus is not on the handling of missing data, we only consider the 301 complete cases. To demonstrate the effect of external borrowing, we simulate hypothetical external control data. 
Our reasons for using simulated external control data (instead of real data) are as follows. First, although many acupuncture trials  have been conducted and published, patient-level data are generally not publicly available. Second,  by generating synthetic data, we can control the degree to which the external and trial data are commensurate. Assume there is a single external data source with 200 patients. We consider the following three scenarios:
\begin{enumerate}[noitemsep,nolistsep,leftmargin=.25in]
\item (Scenario 1) We generate external control data by mimicking the trial control data, resembling a conditional independent case. In particular, we use the \texttt{synthpop} method \citep{nowok2016synthpop}, which allows one to generate a synthetic dataset that preserves the essential statistical features of an observed dataset.
\item (Scenario 2) We oversample patients with low baseline headache scores and undersample patients with high baseline headache scores, although the outcome values are generated using the \texttt{synthpop} method by mimicking similar patients in the trial control data. In other words, the distributions of patient characteristics are different in the trial and external data, although the control outcome is conditionally independent of data source.
\item (Scenario 3) We first generate external control data using the \texttt{synthpop} method, and then update the outcomes by
\begin{align*}
[Y \mid T = 0, \bX, S = 1]  = \tilde{Y} + 5 - 0.05 \cdot \text{baseline headache score},
\end{align*}
where $\tilde{Y}$ and $Y$ stand for the originally generated and updated control outcome values, respectively. In this case, the patient population is similar between the trial and external data, but the control outcome is not conditionally independent of data source.
\end{enumerate}
Figure \ref{fig:real_data} shows the trial data and simulated external control data under each scenario.

We fit the BART model to the dataset under each scenario. For comparison, we also consider a version of BART that does not make use of external data, i.e., the model is fitted to the trial data only. The estimated CATE and its 95\% credible interval in each scenario are reported in Table \ref{tbl:rea_data}.
As expected, when external data are available, the BART model can borrow information to achieve more precise estimates. When the control outcome is not conditionally independent of data source, borrowing may lead to some bias (see estimated CATE for Scenario 3). But since BART borrows information in a judicious manner, the bias is not substantial.

\begin{table}[h!]
\begin{center}
\begin{tabular}{lcccc}
\toprule
\multicolumn{1}{c}{Data}  & \multicolumn{1}{c}{Est. CATE} & \multicolumn{1}{c}{95\% CI} & \multicolumn{1}{c}{CI length} & \multicolumn{1}{c}{$p$-value} \\ \midrule
Trial data only & $4.38$ & $(2.06, 6.70)$ & $4.64$ & -- \\
Trial + external (Scn. 1) & $4.38$ & $(2.14, 6.60)$ & $4.46$ & 0.68  \\
Trial + external (Scn. 2) & $4.38$ & $(2.10, 6.65)$ & $4.55$ & 0.94 \\
Trial + external (Scn. 3) & $4.33$ & $(2.10, 6.57)$ & $4.47$ & 0.02  \\
\bottomrule
\end{tabular}
\end{center}
\caption{Results for the acupuncture trial with simulated external control data.  The control outcome is conditionally independent of the data source indicator under Scenarios 1 and 2 but not under Scenario 3. The $p$-values are obtained from the permutation tests for conditional independence.} 
\label{tbl:rea_data}
\end{table}

An \textit{ad hoc} permutation test can be performed within the BART modeling framework to evaluate the conditional independent assumption of the control outcome and data source \citep{kapelner2016bartmachine}.
Specifically, we may permute the data source indicator,  thereby destroying the relationship between $S$ and $Y$ (given that $\bX$ is in the model), fit a new BART model from this permuted design matrix and record how the new model fits the data. The BART model fit can be characterized by the pseudo-$R^2$: $1 - \sum_i(y_i - \hat{y}_i)^2 / \sum_i(y_i - \bar{y})^2$, where $\hat{y}_i$ denotes the fitted value of $y_i$. 
Repeating this permutation procedure many times, we obtain a null distribution of pseudo-$R^2$'s.
The $p$-value can then be defined as the proportion of null pseudo-$R^2$'s greater than the pseudo-$R^2$ of the BART fit to the original dataset.
This is a one-sided test. If the $p$-value is small, it is likely that the data source indicator $S$ is an important predictor for $Y$, after controlling for $\bX$.
This permutation test can be easily done using the \texttt{cov\_importance\_test} function in the R package \texttt{bartMachine} \citep{kapelner2016bartmachine}.
The $p$-value for the permutation test (based on 100 permutation samples) under each scenario is reported in Table \ref{tbl:rea_data}. From the results, we can see that $S$ has a strong effect on $Y$ under Scenario 3 but not under Scenarios 1 and 2. This is consistent with the simulation truth.

\section{Discussion}
\label{sec:discussion}

We have explored the capacity of BART to incorporate external data into the analysis of clinical trials. BART adaptively pools information across data sources to improve the precision of treatment effect estimates. Simulation studies have shown that the proposed BART method compares favorably to a hierarchical linear model and a normal-normal hierarchical model. We have focused on continuous outcomes, but the \texttt{BART} package also supports binary and time-to-event outcomes, making it easy to extend the proposed method.

Other flexible regression methods, such as Gaussian process regression (GPR, \citealp{rasmussen2006gaussian}), are also suitable for our application. 
However, we have chosen BART for its several attractive features outlined in Section \ref{sec:intro}. Moreover, empirically we have found that BART is computationally efficient, while GPR suffers from a cubic time complexity and finds it hard to handle a large number of covariates.

We have followed the default model specification in \cite{chipman2010bart}. An interesting future direction is to further tailor the BART model specifically for our application.
For example, the default BART model assumes a uniform prior on the splitting variable assignment. An alternative prior can be specified to discourage (or encourage) selecting the data source indicator as a splitting variable, achieving stronger (or weaker) borrowing across data sources.
We have included the data source indicator as a categorical covariate in the BART model (see Equation \ref{eq:model_ctl}). An alternative model specification \citep{tan2019bayesian}  is
\begin{align*}
[Y \mid T = 0, \bX = \bx, S = s] \sim \N \left[ f_0 (\bx) + h_0(\bx, s), \sigma_0^2 \right],
\end{align*}
where $f_0 (\bx) = \sum_{j = 1}^{m_0} g(\bx; \T_{0j}, \M_{0j})$ is a sum-of-trees model, and $h_0(\bx, s)$ is a parametric model, for example, $h_0(\bx, s) = \alpha_{0s}$ or $h_0(\bx, s) = \alpha_{0s} + \bx^{\top} \bbeta_{0s}$. By tuning the prior parameters for $\alpha_{0s}$ and $\bbeta_{0s}$, one can also control the degree of borrowing across data sources.
Finally, for simplicity, the error variance ($\sigma_0^2$) has been assumed common across data sources, but this assumption may be relaxed.

\bibliographystyle{apalike}
\bibliography{ref_hc}

\end{document}